<

# A Comparative Tutorial of the Histogram-based Image Segmentation Methods

Zhenzhou Wang*

School of computer science and technology, Huaibei Normal University, Huaibei City, Anhui Province, 235000, China

*Corresponding author
*E-mail:* wangzz@chnu.edu.cn

## Abstract

The histogram of an image is the accurate graphical representation of the numerical grayscale distribution and it is also an estimate of the probability distribution of image pixels. Therefore, histogram has been widely adopted to calculate the clustering means and partitioning thresholds for image segmentation. There have been many classical histogram-based image segmentation methods proposed and played important roles in both academics and industry. In this tutorial, the histories and recent advances of the histogram-based image segmentation techniques are first reviewed and then they are divided into four categories: (1) the means-based method, (2) the Gaussian-mixture-model-based method, (3) the entropy-based method and (4) the feature-points-based method. The purpose of this tutorial is threefold: 1) to teach the principles of the classical histogram-based image segmentation methods to the interested readers; 2) to evaluate the advantages and disadvantages of these classical histogram-based image segmentation methods objectively; 3) to compare the performances of these classical histogram-based image segmentation methods with state-of-the-art deep learning based methods objectively.

## 1. Introduction

Image segmentation divides a grayscale or color image into different meaningful regions and extracts the target objects, which plays an important role in many fields, such as real-time 3D measurement [1-3] and smart medical care [4-5]. Ever since 1960s, image segmentation studies have received significant attention from different researchers around the world and the histogram-based image segmentation methods have also made great progress in the following decades. For instance, the k-means method [6] had been proposed in the 1960s and used for image segmentation [7] in the 1970s. Both the expectation maximization (EM) method [8] and the maximum inter-class difference method [9] were proposed for segmenting the grayscale images in the 1970s. The fuzzy C-means method [10], the maximum-entropy method [11-12], the fuzzy-entropy method [13], the cross-entropy method [14] and the feature-points-based method [15] were proposed for image segmentation in the 1980s and 1990s. Due to the great variety of the images acquired in different application scenarios, the performances of these image segmentation methods usually vary significantly in different applications. As a result, these methods could be used directly in some applications or need to be optimized extensively to improve their performances in other applications.

Based on the working principles, there are mainly four categories of histogram-based image segmentation methods. The first category is the means-based methods that include the k-means methods [7,16-17], the Otsu's method [9, 18-19] and the fuzzy C-means methods [10,20-22]. The similarity of these methods is that all of them compute the class means iteratively. The differences of these methods are that the K-means method only calculates the class means in each iteration, the Otsu method calculates a proportion coefficient for each class mean, and the fuzzy C-means method calculates a fuzzy membership function for each class mean. Despite the straightforward working principles, researchers are still working on optimizing the performances of these methods nowadays. A novel density-based technique to accurately identify cluster cores was used to improve the K-means clustering accuracy in [16] and better initialization algorithms were also used to improve the K-means clustering accuracy in [17]. To optimize the performance of the Otsu's method, a support vector machine that is based on the Levy horse was proposed to compute multi-level thresholds optimally with significant improvement in [18]. In addition, the chameleon swarm algorithm was improved to increase the convergence of the Otsu's method in [19]. To improve the accuracy of the fuzzy C-means method, the gradient descent was used to solve an unconstrained fuzzy C-means algorithm and accordingly a novel deep fuzzy clustering model was obtained for clustering [20]. The new possibilistic fuzzy weighted local information factor was introduced into the interval type-2 possibilistic fuzzy C-means for noisy image segmentation [21] and the fuzzy membership function with Shannon's entropy-based variation was used for fuzzy C-means clustering based image segmentation [22].

The second category is the Gaussian-mixture-model-based method that uses the EM methods [8, 23] to calculate the mixture model's parameters. This category of method assumes that the histogram distribution consists of $K$ Gaussian distributions and the histogram distribution could thus be denoted as a Gaussian distribution mixture model. The maximum A posteriori method [8], the least squared error method and the cross-correlation method could be used to generate the cost function for minimization by iteration. The iteration processes of the above three methods are the same, the iteration times are usually different and the accuracies vary insignificantly. In recent years, the spatial information was incorporated into the Gaussian mixture model to improve the image segmentation accuracy [23].

The third category is the entropy-based methods that include the max Shannon entropy methods [11-12,24], the fuzzy Shannon entropy methods [13,25], and the Shannon cross-entropy methods [14, 26-27]. The max Shannon entropy methods have been proved to be similar to the maximum-likelihood and Bayesian methods [11]. In histogram-based threshold selection, the max Shannon entropy method globally finds the thresholds that make the Shannon information entropy of all the independent pixel classes in the histogram distribution maximum. Similar to the max Shannon entropy method, the fuzzy Shannon entropy method also globally finds the thresholds that make the Shannon information entropy of all the independent pixel classes in the histogram distribution maximum. Different from the max Shannon entropy method, the fuzzy entropy method adds a fuzzy membership function to its cost function. The cross Shannon entropy calculates the difference between two probability distributions. It had become a well-known concept of information theory for many decades and had been widely used in many different fields. The cross Shannon entropy method uses the cross-entropy distance or the metric cross entropy distance to define the cost function. In recent years, an adaptive entropy fusion method was proposed to improve the segmentation accuracy in different lighting scenes [24]. The fuzzy entropy multi-level thresholding was used for fundus image modalities segmentation [25]. The human mental search algorithm [26] and the chimp-inspired remora optimization algorithm [27] were used respectively to minimize the cross-entropy for efficient multi-level thresholding.

The fourth category is the feature-points-based method that uses the grayscale positions that generate the feature points with global or local maximum rates of change to calculate the clustering means or partitioning thresholds [15, 28-29]. This category of method has been based on the assumption that the rate of change at the border of two pixel-classes in the histogram distribution will reach the global maximum or at least a local maximum and the rate of change at the center of a pixel-class will also reach the global maximum or at least a local maximum, which can be easily observed. In addition, the points with the global or local maximum rate of change can be treated as the histogram feature points. For instance, the representative histogram feature points are the histogram peaks and histogram valleys. Since the shape of the histogram could reflect the histogram feature points, this category of method is also called histogram-shape-based method in [30]. In recent years, the slope difference distribution has been proposed to compute the thresholds or the clustering centers for the feature-points-based method to achieve robust image segmentation [28-29].

Although these histogram-based methods are still extensively researched or used in recent years [31-32], they have the bottleneck problem of segmenting images with non-uniform and complex backgrounds. Since the deep learning methods extract the objects based on their features instead of the gray levels, they could overcome this bottleneck problem potentially. As a result, deep learning has been paid a lot of attention in recent years and many large-scale annotation-based image datasets have emerged for training the deep neural networks. Although deep learning has made great achievements in image segmentation, it still has many challenges [33-34]. For instance, the annotations are usually sparse and limited. The classes are usually imbalanced. The overfitting problem and the gradient vanishing problem may exist. The training time is usually considerably long. In addition, some applications might have very limited resources to train the deep learning methods. At this time, the histogram-based image segmentation methods are usually preferred as a better choice.

The organization of this tutorial is as follows. Firstly, the principles of the means-based methods that include the K-means method, the Otsu's method and the fuzzy C-means method are described. Secondly, the principles of the Gaussian-mixture-model-based methods that include the maximum A posteriori method, the least squared error method and the cross-correlation method are described. Thirdly, the principles of the entropy-based methods that include the maximum entropy method, the fuzzy entropy method and the cross-entropy method are described. Fourthly, the principles of the feature-points-based methods that include the histogram-peak-and valley method and the slope-difference-distribution method are described. Fifthly, these described methods are evaluated and compared with the deep learning methods on the synthesized images. Sixthly, these described methods are evaluated and compared with the deep learning methods on the real images. At last, conclusions are drawn.

## 2. The Means-based Method

### *2.1 The K-means method*

The popular K-means method [7] updates the means of the histogram iteratively to find the minimum of the summed squared errors between the clustered pixels and the means they belong to, which could be formulated as:

$$J^2 = \sum_{i=1}^{T_1} (i-\mu_1)^2 + \sum_{k=2}^{K-1} \sum_{i=T_{k-1}+1}^{T_k} (i-\mu_k)^2 + \sum_{i=T_{K-1}+1}^{N} (i-\mu_K)^2 \quad (1)$$

$\mu_k$ represents the mean of the $k$th class and it is computed by the following equation.

$$\mu_k = \begin{cases} \sum_{i=1}^{T_1} \dfrac{i \times p(y_i)}{\sum_{i=1}^{T_1} p(y_i)}; k=1 \\ \sum_{i=T_{k-1}+1}^{T_k} \dfrac{i \times p(y_i)}{\sum_{i=T_{k-1}+1}^{T_k} p(y_i)}; k=2,...,K-1 \\ \sum_{i=T_{K-1}+1}^{N} \dfrac{i \times p(y_i)}{\sum_{i=T_{K-1}+1}^{N} p(y_i)}; k=K \end{cases} \quad (2)$$

In each iteration, the pixels are classified again based on their distances to the clustering center $\mu_k$ with the following equation.

$$\bar{k} = \arg\min_{k=1,...,K} |i-\mu_k|; i=1,2,...,N \quad (3)$$

$\bar{k}$ denotes the class number that the pixel $i$ belongs to. Each pixel $y_i$ will belong to one of the $K$ pixel classes that are formulated as:

$$y_i^k = \begin{cases} [1,\dots,T_1]; k=1 \\ [T_{k-1}+1,\dots,T_k]; k=2,\dots,K-1 \\ [T_{K-1}+1,\dots,N]; k=K \end{cases} \quad (4)$$

With the newly determined pixel classes, $\mu_k$ are computed again by Eq. (2). The iteration stops until the cost function formulated by Eq. (1) converges, i.e.,

$$J^{n+1} - J^n < \Delta \quad (5)$$

where $J^{n+1}$ denotes the cost function formulated by Eq. (1) in the next iteration and $J^n$ denotes the cost function formulated by Eq. (1) in the current iteration. $\Delta$ denotes a predefined threshold. Or the means vector $(\mu_1, \mu_2, \dots, \mu_K)$ converges, i.e.,

$$(\mu_1, \mu_2, \dots, \mu_K)^{n+1} - (\mu_1, \mu_2, \dots, \mu_K)^n < \Delta \quad (6)$$

The thresholds could also be computed as follows.

$$T_k = \frac{\mu_k + \mu_{k+1}}{2}; k = 1, 2, \dots K-1 \quad (7)$$

*2.2 The Otsu method*

The famous Otsu method [9] globally or iteratively finds the means that maximize the differences between the interclasses in the histogram distribution, which can be formulated as follows.

$$J = \sqrt{\sum_{k=1}^{K} w_k (\mu - \mu_k)^2} \quad (8)$$

$\mu$ denotes the mean of the whole histogram distribution and it is computed with the following equation.

$$\mu = \sum_{i=1}^{N} \frac{i \times p(y_i)}{\sum_{i=1}^{N} p(y_i)} \quad (9)$$

$w_k$ denotes the proportion of the $k$th class in the histogram distribution and it is computed with the following equation.

$$w_k = \begin{cases} \sum_{i=1}^{T_1} p(y_i); k=1 \\ \sum_{i=T_{k-1}+1}^{T_k} p(y_i); k=2,\dots,K-1 \\ \sum_{i=T_{K-1}+1}^{N} p(y_i); k=K \end{cases} \quad (10)$$

$\mu_k$ represents the mean of the $k$th class and it is computed by the following equation.

$$\mu_k = \begin{cases} \sum_{i=1}^{T_1} \frac{i \times p(y_i)}{w_1}; k=1 \\ \sum_{i=T_{k-1}+1}^{T_k} \frac{i \times p(y_i)}{w_k}; k=2,\dots,K-1 \\ \sum_{i=T_{K-1}+1}^{N} \frac{i \times p(y_i)}{w_K}; k=K \end{cases} \quad (11)$$

In each iteration, the pixels are classified again based on their distances to the clustering center $\mu_k$ with Eqs. (3-4). With the newly determined pixel classes, $w_k$ and $\mu_k$ are computed again by Eq. (10) and Eq. (11) respectively. The iteration stops until the cost function formulated by Eq. (8) converges or the means vector $(\mu_1, \mu_2, \dots, \mu_K)$ converges.

*2.3 The fuzzy-C-means method*

Like the K-means method, the popular fuzzy-C-means method [10] also updates the means of the histogram iteratively to find the minimum of the summed squared errors between the clustered pixels and the means they belong to. Different from the K-means method, it adds a fuzzy membership function to its cost function and it could be formulated as:

$$J^2 = \sum_{i=1}^{T_1} u_{i1}^m (i - \mu_1)^2 + \sum_{k=2}^{C-1} \sum_{i=T_{k-1}+1}^{T_k} u_{ik}^m (i - \mu_k)^2 + \sum_{i=T_{K-1}+1}^{N} u_{iC}^m (i - \mu_C)^2 \quad (12)$$

$u_{ij}$ denotes the fuzzy membership function and it is computed with the following equation.

$$u_{ij} = \frac{1}{(x_i - \mu_j)^{\frac{2}{m-1}} \sum_{k=1}^{C} \left( \frac{1}{(x_i - \mu_k)^{\frac{2}{m-1}}} \right)} \quad (13)$$

$\mu_j$ denotes the mean of the $j$th class in the histogram distribution and it is computed with the following equation.

$$\mu_j = \frac{\sum_{i=1}^{N} i \times u_{ij}^m \times p(y_i)}{\sum_{i=1}^{N} u_{ij}^m \times p(y_i)}; j = 1, \dots, C \quad (14)$$

In each iteration, the fuzzy membership $u_{ij}$ of each pixel is computed again with Eq. (13). With the newly computed fuzzy membership $u_{ij}$, $\mu_j$ is computed again by Eq. (14). The iteration stops until the cost function formulated by Eq. (12) converges or the means vector $(\mu_1, \mu_2, \dots, \mu_C)$ converges. The thresholds are then computed by Eq. (7).

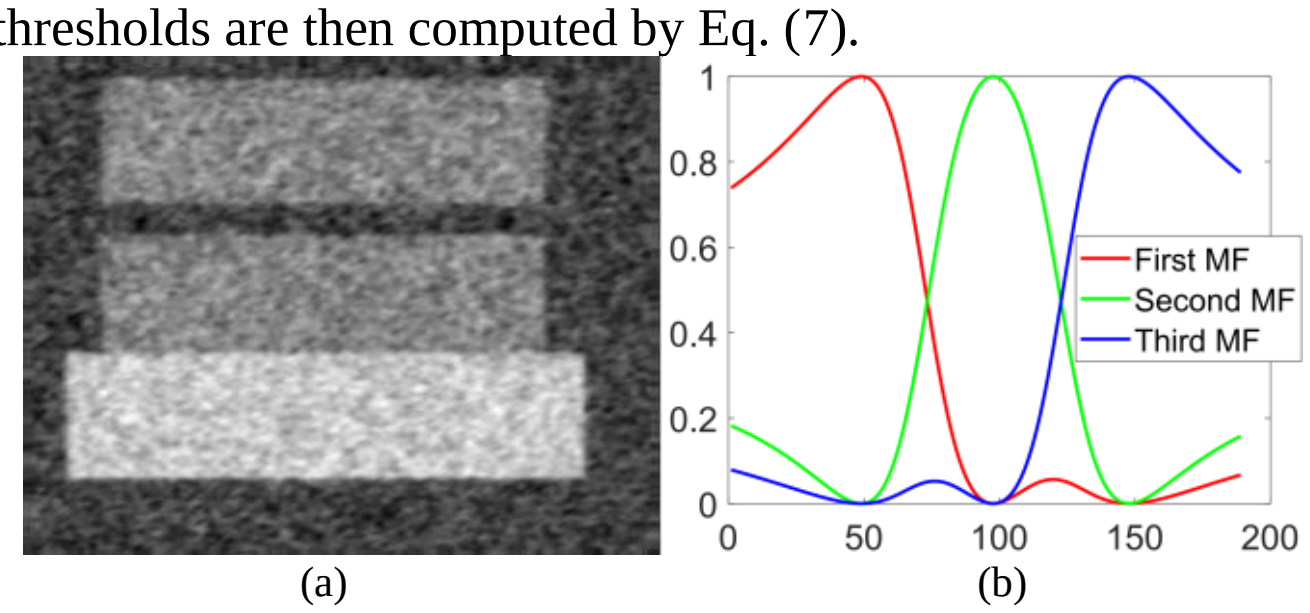

Fig. 1. Demonstration of the synthesized image and its computed fuzzy membership functions. (a) The synthesized image; (b) The computed fuzzy membership functions.

An image with three rectangles is synthesized by MATLAB as follows. The gray levels of the background belong to the first pixel class and are assigned as 100. The gray levels of the two smaller rectangles belong to the second pixel class and are assigned as 150. The gray levels of the largest rectangle belong

<

to the third pixel class and are assigned as 200. Then the Gaussian noises are added by multiplying the standard normal distribution with an amplitude 50. Fig. 1 (a) shows the synthesized image. The computed fuzzy membership functions (MF) by Eq. (13) for the first, second and third pixel class are demonstrated as the red, green and blue curve respectively. They are denoted as "First MF", "Second MF" and "Third MF" respectively in Fig. 1 (b).

### 3. The Gaussian-mixture-model-based Method

This type of method assumes that the histogram distribution is composed of $K$ Gaussian distributions and the histogram distribution could thus be denoted as a Gaussian distribution mixture model. Suppose the $k$th Gaussian distribution has the parameter that is denoted as $\phi_k = \left(\mu_k, \sigma_k^2\right), k = 1,...,K$ , then the $k$th conditional Gaussian distribution given the observed histogram distribution $p\left(y_i\right), y_i = i = 1,2,...,N$ is formulated as:

$$p\left(x_i \mid y_i, \phi_k\right) = \frac{1}{\sigma_k\sqrt{2\pi}} e^{-\frac{\left(x_i-\mu_k\right)^2}{2\sigma_k^2}} ; i = 1,...,N; k = 1,...,K \quad (15)$$

$g_1$ denotes the smallest grayscale value and $g_2$ denotes the largest grayscale value. Accordingly, the Gaussian mixture model of the histogram is formulated as:

$$p\left(x_i \mid y_i, \phi\right) = \sum_{k=1}^{K} p\left(x_i \mid y_i, \phi_k\right); i = 1,2,...,N \quad (16)$$

*3.1 The maximum A posteriori method*

The maximum A posteriori (MAP) method is a way of computing the maximum likelihood (ML) for the posterior probability of the histogram distribution that is formulated as:

$$p\left(y_i \mid x_i, \phi\right) = \frac{p\left(y_i\right)\sum_{k=1}^{K} p\left(x_i \mid y_i, \phi_k\right)}{\sum_{i=1}^{N}\left(p\left(y_i\right)\sum_{k=1}^{K} p\left(x_i \mid y_i, \phi_k\right)\right)}; i = 1,2,...,N \quad (17)$$

The denominator part of the above equation is a constant. Thus, to maximize the posterior probability of the histogram distribution is to maximize the numerator part of the above equation, i.e.,

$$p\left(y_i \mid x_i, \phi\right) \propto p\left(y_i\right)\sum_{k=1}^{K} p\left(x_i \mid y_i, \phi_k\right); i = 1,2,...,N \quad (18)$$

The MAP likelihood function is thus formulated as:

$$L\left(\phi\right) = \prod_{i=1}^{N}\left(p\left(y_i\right)\sum_{k=1}^{K} p\left(x_i \mid y_i, \phi_k\right)\right) \quad (19)$$

The logarithmic function is used to avoid the above equation approaching 0 infinitely. Accordingly, the logarithmic MAP likelihood function is formulated as:

$$J = \log\left(L\left(\phi\right)\right) = \sum_{i=1}^{N} p\left(y_i\right)\log\left(\sum_{k=1}^{K} p\left(x_i \mid y_i, \phi_k\right)\right) \quad (20)$$

The expectation maximization (EM) method is a systematic method of estimating parameters from incomplete data [8] and it provides a way of determining the ML estimate of the parameters $\phi_k = \left(\mu_k, \sigma_k^2\right), k = 1,...,K$ by updating them iteratively with the following equations.

$$\mu_k = \sum_{i=1}^{N} \frac{i \times p\left(y_i\right) \times p\left(x_i \mid y_i, \phi_k\right) / p\left(x_i \mid y_i, \phi\right)}{\sum_{i=1}^{N} p\left(y_i\right) \times p\left(x_i \mid y_i, \phi_k\right) / p\left(x_i \mid y_i, \phi\right)} \quad (21)$$

$$\sigma_k^2 = \sum_{i=1}^{N} \frac{\left(i-\mu_k\right)^2 \times p\left(y_i\right) \times p\left(x_i \mid y_i, \phi_k\right) / p\left(x_i \mid y_i, \phi\right)}{\sum_{i=1}^{N} p\left(y_i\right) \times p\left(x_i \mid y_i, \phi_k\right) / p\left(x_i \mid y_i, \phi\right)} \quad (22)$$

The EM method exits the iteration until the logarithmic MAP likelihood function converges. After $\phi_k = \left(\mu_k, \sigma_k^2\right), k = 1,...,K$ is determined, the thresholds are computed by Eq. (7). This method is named as EM-MAP in this study.

*3.2 The least squared error method*

The least squared error method computes the minimum of the root squared errors between the histogram distribution $p\left(y_i\right)$ and the Gaussian distribution mixture model $p\left(x_i \mid y_i, \phi\right)$ that is formulated by the following equation.

$$J = \sqrt{\sum_{i=1}^{N}\left(p\left(y_i\right) - p\left(x_i \mid y_i, \phi\right)\right)^2} \quad (23)$$

In the same way, the EM method is used to determine the ML estimate of the parameters $\phi_k = \left(\mu_k, \sigma_k^2\right), k = 1,...,K$ by updating them in an iterative way by Eq. (21) and Eq. (22). Then, the iteration is terminated by Eq. (5) or Eq. (6) and the thresholds are computed by Eq. (7). This method is named as EM-LSE in this study.

*3.3 The cross-correlation method*

If we treat the histogram and the Gaussian mixture model as two continuous signals, $h\left(t\right)$ and $g\left(t\right)$ respectively, their cross-correlation function is formulated as:

$$R_{hg}\left(\tau\right) = \int_{-\infty}^{+\infty} h\left(t\right) g\left(t+\tau\right) dt \quad (24)$$

For the discrete formulations, the histogram is represented as $p\left(y_i\right)$ and Gaussian mixture model is represented as $p\left(x_i \mid y_i, \phi\right)$ . Thus, the cross-correlation function can be formulated as:

$$J = R_{hg}\left(0\right) = \sum_{i=1}^{N} p\left(y_i\right) p\left(x_i \mid y_i, \phi\right) \quad (25)$$

According to the cross-correlation theory, $R_{hg}\left(0\right)$ reaches the maximum value when the histogram and the Gaussian mixture model are most similar. Therefore, Eq. (25) could also be used as the cost function by the EM method to exit the iteration defined by Eq. (5) and Eq. (7) is used to compute the thresholds. This method is named as EM-CC in this study.

The iteration processes of the above three methods are the same, the iteration times are usually different and the accuracies vary insignificantly. The synthesized image shown in Fig. 1 (a) is used to demonstrate the iteration process of the EM-MAP method and the iteration results at different times are shown in Fig. 2 (a), (b) (c) and (d) respectively. Where "OH-D" represents the histogram distribution of the synthesized image and "GMM-D" represents the Gaussian mixture model distribution formulated by Eq. (16).

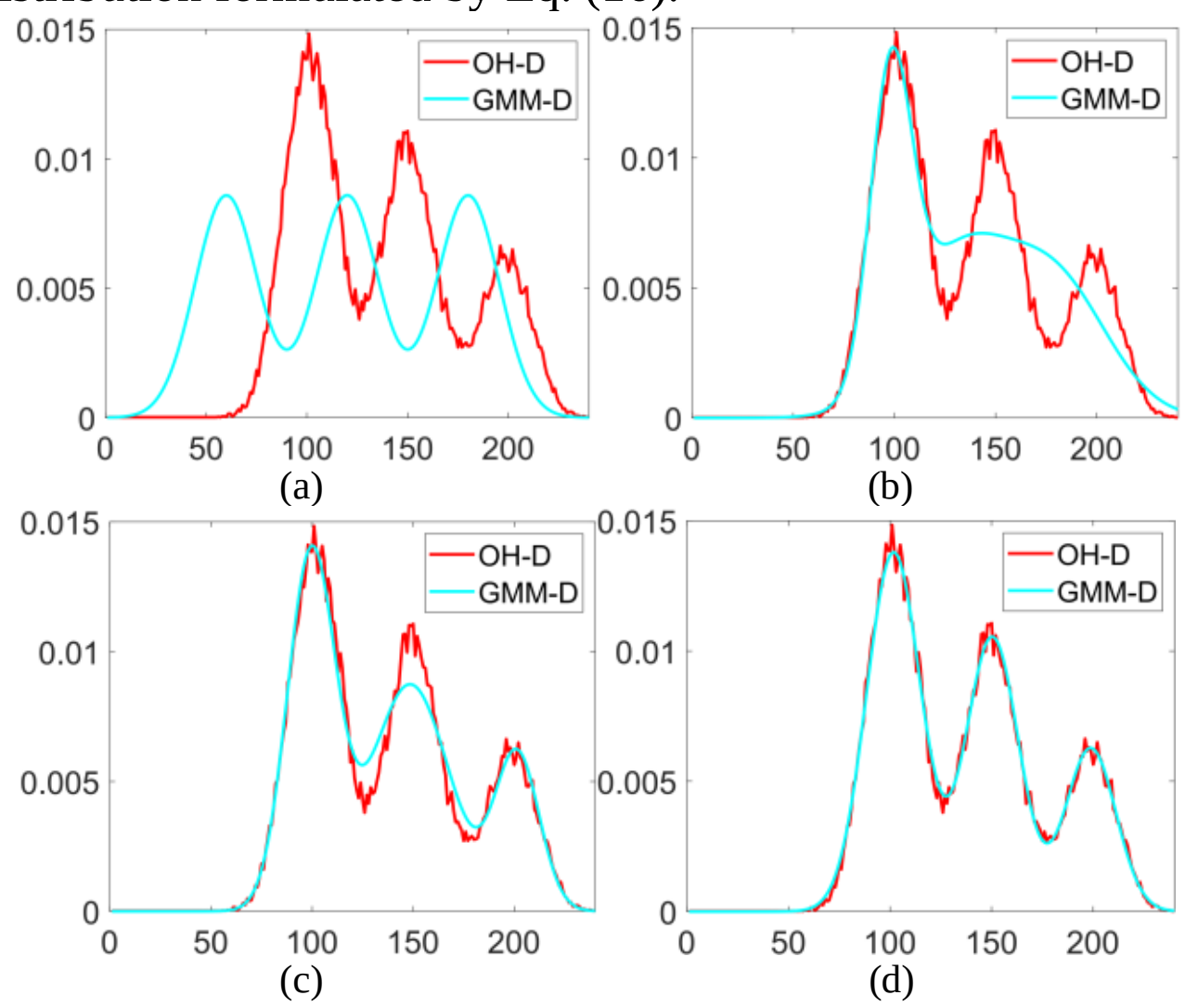

Fig. 2. Demonstration of the iteration process of EM. (a) The results of the 1th iteration; (b) The results of the 20th iteration; (c) The results of the 60th iteration; (d) The results of the 96th iteration.

## 4. The Entropy-based Method

### *4.1 The max-entropy-based method*

Shannon information entropy is able to quantify the information and represent its uncertainty. In other words, the entropy of the information is proportional to its uncertainty. The Shannon information entropy of the histogram distribution can be formulated as follows.

$$H(y_i) = -\sum_{i=1}^{N} p(y_i)\log\left(p(y_i)\right) \tag{26}$$

The max Shannon entropy methods have been proved to be similar to the maximum-likelihood and Bayesian methods [11]. The probability distributions $p_k(y_i); k=1,2,\dots,K$ of the independent pixel classes in the histogram distribution are computed as follows.

$$p_k(y_i) = \begin{cases} \dfrac{p(y_i)}{\sum_{i=1}^{T_1} p(y_i)}; k=1; 1 \le i \le T_1 \\ \dfrac{p(y_i)}{\sum_{i=T_{k-1}+1}^{T_k} p(y_i)}; k=2,\dots,K-1; T_{k-1}+1 \le i \le T_k \\ \dfrac{i \times p(y_i)}{\sum_{i=T_{K-1}+1}^{N} p(y_i)}; k=K; T_{K-1}+1 \le i \le N \end{cases} \tag{27}$$

In histogram-based threshold selection, the max entropy method globally finds the thresholds that make the Shannon information entropy of all the independent pixel classes in the histogram distribution maximum, which can be formulated as:

$$\begin{aligned} J = &-\sum_{i=1}^{T_1} p_1(y_i)\log\left(p_1(y_i)\right) \\ &-\sum_{k=2}^{K-1}\sum_{i=T_{k-1}+1}^{T_k} p_k(y_i)\log\left(p_k(y_i)\right) - \sum_{i=T_{K-1}+1}^{N} p_K(y_i)\log\left(p_K(y_i)\right) \end{aligned} \tag{28}$$

The thresholds $\{T_1, T_2, \dots, T_{K-1}\}$ that makes Eq. (28) maximum could be found by a global search.

### *4.2 The fuzzy-entropy-based method*

Like the max entropy method, the fuzzy entropy method [13] also globally finds the thresholds that make the information entropy of all the independent pixel classes in the histogram distribution maximum. Unlike the max entropy method, the fuzzy entropy method adds a fuzzy membership function to its cost function that could be formulated as:

$$\begin{aligned} J = &-\sum_{i=1}^{T_1} m_1 p_1(y_i)\log\left(p_1(y_i)\right) \\ &-\sum_{k=2}^{K-1}\sum_{i=T_{k-1}+1}^{T_k} m_k p_k(y_i)\log\left(p_k(y_i)\right) - \sum_{i=T_{K-1}+1}^{N} m_K p_K(y_i)\log\left(p_K(y_i)\right) \end{aligned} \tag{29}$$

$m_k$ denotes the fuzzy membership function and it is computed as follows.

$$m_k = \begin{cases} \left(1 + \dfrac{\left|i - \sum_{i=1}^{T_1} i \times p(y_i)\right|}{N}\right)^{-1}; & k=1;\ i=1,\dots,T_1 \\ \left(1 + \dfrac{\left|i - \sum_{i=T_{k-1}+1}^{T_k} i \times p(y_i)\right|}{N}\right)^{-1}; & k=2,\dots,K-1;\ i=T_{k-1}+1,\dots,T_k \\ \left(1 + \dfrac{\left|i - \sum_{i=T_{K-1}+1}^{N} i \times p(y_i)\right|}{N}\right)^{-1}; & k=K;\ i=T_{K-1}+1,\dots,N \end{cases} \tag{30}$$

The thresholds $\{T_1, T_2, \dots, T_{K-1}\}$ that makes Eq. (29) maximum could be found by a global search.

### *4.3 The cross-entropy-based method*

Cross entropy is able to measure the difference between two probability distributions, which makes it important in information theory. The cross-entropy distance between a posteriori probability distribution $q(y_i)$ and its priori distribution $p(y_i)$ can be formulated as follows.

$$H_{CE}(q,p)=\sum_{i=1}^{N} q(y_i)\log\left(\frac{q(y_i)}{p(y_i)}\right), i=1,\dots,N \qquad (31)$$

The metric cross entropy distance can be formulated as follows.

$$H_m(q,p)=H_{CE}(q,p)+H_{CE}(p,q)= \\ \sum_{i=1}^{N} q(y_i)\log\left(\frac{q(y_i)}{p(y_i)}\right)+\sum_{i=1}^{N} p(y_i)\log\left(\frac{p(y_i)}{q(y_i)}\right) \qquad (32)$$

Based on the cross entropy distance, the cost function can be formulated as follows [14].

$$J=\sum_{i=1}^{T_1} p(y_i)\mu_1\log\left(\frac{\mu_1}{i}\right)+ \\ \sum_{k=2}^{K-1}\sum_{i=T_{k-1}+1}^{T_k} p(y_i)\mu_k\log\left(\frac{\mu_k}{i}\right)+\sum_{i=T_{K-1}+1}^{N} p(y_i)\mu_K\log\left(\frac{\mu_K}{i}\right) \qquad (33)$$

Based on the metric cross entropy distance, the cost function can be formulated as follows [78].

$$J=\sum_{i=1}^{T_1} p(y_i)\left(\mu_1\log\left(\frac{\mu_1}{i}\right)+i\log\left(\frac{i}{\mu_1}\right)\right) \\ +\sum_{k=2}^{K-1}\sum_{i=T_{k-1}+1}^{T_k} p(y_i)\left(\mu_k\log\left(\frac{\mu_k}{i}\right)+i\log\left(\frac{i}{\mu_k}\right)\right) \\ +\sum_{i=T_{K-1}+1}^{N} p(y_i)\left(\mu_K\log\left(\frac{\mu_K}{i}\right)+i\log\left(\frac{i}{\mu_K}\right)\right) \qquad (34)$$

The thresholds $\{T_1,T_2,\dots,T_{K-1}\}$ that makes Eq. (33) or (34) maximum could be found by a global search.

## 5. The Feature-points-based Method

It is observed that the rate of change at the border of two adjacent pixel classes in a histogram distribution will reach the global maximum or a local maximum. In addition, the rate of change at the center of a pixel-class will also reach the global maximum or a local maximum. On the other hand, the points that yield the global or local maximum rate of change can be treated as the histogram feature points. Thus, the grayscale positions with global or local maximum rates of change are used by the histogram-feature-points-based methods as candidates for computing the clustering means and partitioning thresholds. To avoid too many histogram feature points are detected by the histogram-feature-points-based methods, a histogram is usually smoothed beforehand. Therefore, the histogram needs to be filtered before the histogram-feature-points-based methods are used to calculate the clustering means and partitioning thresholds.

### *5.1 Discrete-Fourier-transform-based filtering*

The discrete Fourier transform based filter was proved to be the most effective low-pass filter for histogram smoothing [28-29]. In this study, the histogram distribution $p(y_i), i=1,2,\dots,N$ is rewritten as $p(x), x=1,2,\dots,N$. Accordingly, the discrete Fourier transformation $X(k)$ of the histogram distribution $p(x)$ can be formulated as follows.

$$X(k)=\sum_{x=1}^{N} p(x)e^{\frac{-j2\pi(k-1)(x-1)}{N}}; k=1,2,\dots,N \qquad (35)$$

The inverse discrete Fourier transform is formulated as follows.

$$p(x)=\frac{1}{N}\sum_{k=1}^{N} X(k)e^{\frac{j2\pi(k-1)(x-1)}{N}}; x=1,2,\dots,N \qquad (36)$$

Suppose the band width of the discrete Fourier transform filter is $W$, then the filtered and undistorted histogram $p'(x)$ could be formulated as follows.

$$p'(x)=\frac{1}{N}\sum_{k=1}^{W} X(k)e^{\frac{j2\pi(k-1)(x-1)}{N}} \\ +\frac{1}{N}\sum_{k=N-W+1}^{N} X(k)e^{\frac{j2\pi(k-1)(x-1)}{N}}; x=1,2,\dots,N \qquad (37)$$

### *5.2 The histogram-peak-and-valley-based method*

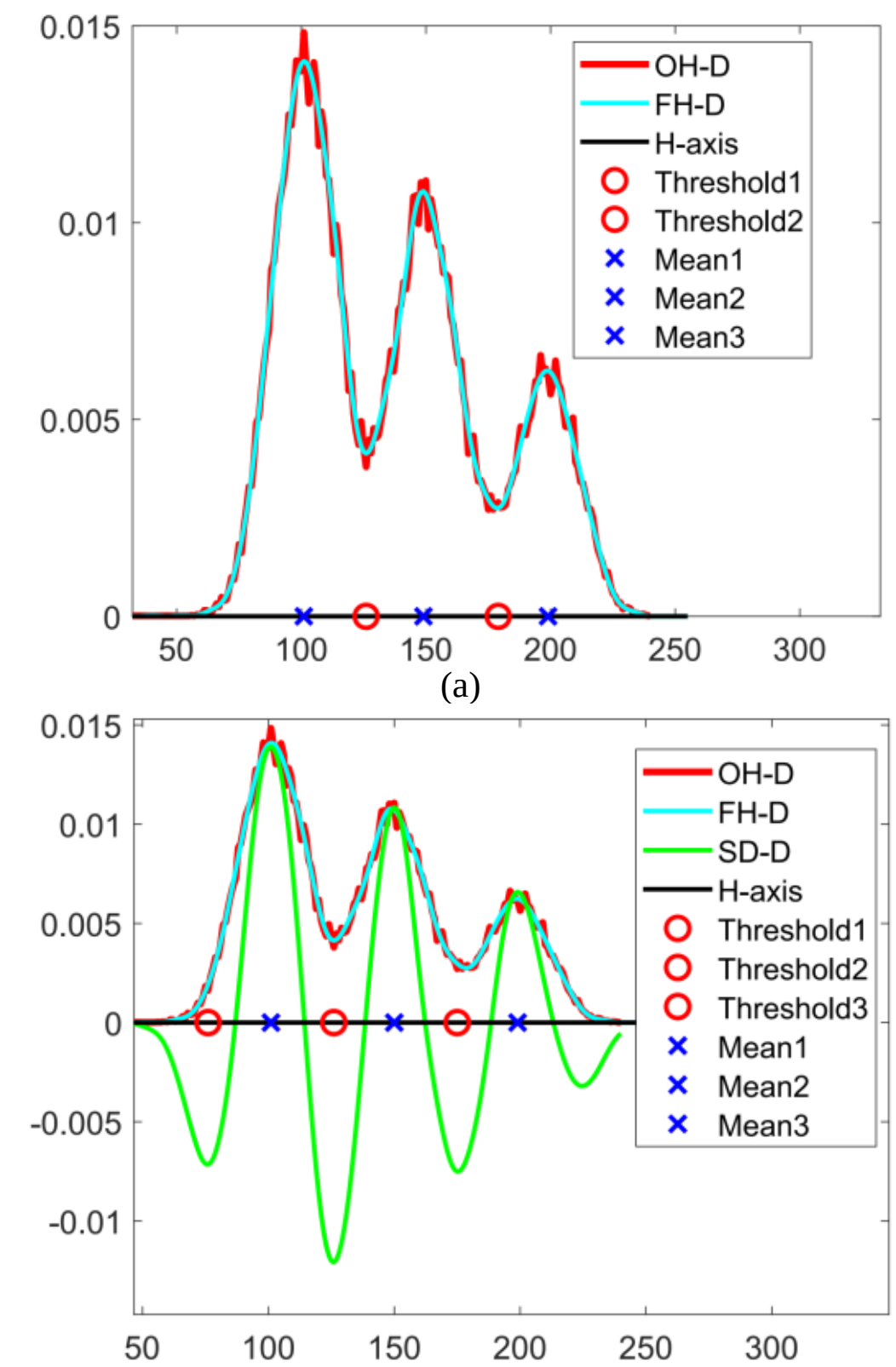

Fig. 3. Demonstration of the rate of change methods. (a) Demonstration of the HPV method; (b) Demonstration of the SDD method.

When the histogram distribution could be denoted as a Gaussian distribution mixture model, the histogram peak feature points correspond to the mean values of the Gaussian distributed pixel classes and the histogram valley feature points correspond to the thresholds to partition the adjacent Gaussian distributed pixel classes. Suppose the detected histogram peak feature points are denoted as $P_k; k=1,2,\dots,K$, the mean of the $k$th pixel class in the histogram distribution is computed as:

$$\mu_k=\{y_i \mid p(y_i)=P_k, i=1,2,\dots,N\} \qquad (38)$$

Suppose the detected histogram valley feature points are denoted as $V_k; k=1,2,\dots,K-1$, the thresholds are computed as:

$$T_k = \{ y_i \mid p(y_i) = V_k, i = 1,2,...,N \} \quad (39)$$

Besides the equation formulated above, the thresholds could also be computed by Eq. (7). This method is named as HPV in this study.

The synthesized image shown in Fig. 1 (a) is used to demonstrate the histogram peak and valley (HPV) based method in Fig. 3 (a). Where "OH-D" denotes the original histogram distribution of the synthesized image, "FH-D" denotes the filtered histogram distribution formulated by Eq. (36) and "H-axis" denotes the horizontal axis. The grayscale values that correspond to the detected valleys are marked by the red circles and they are the candidate thresholds. In this demo, there are two thresholds, Threshold1 and Threshold2. The grayscales that are corresponding to the detected peaks are marked by the blue crosses and they represent the means, Mean1, Mean2 and Mean3 of the three pixel-classes.

*5.3 The slope-difference-distribution-based method*

When the histogram distribution could not be denoted as a Gaussian distribution mixture model, the histogram peak feature points or valley feature points may not exist. In these situations, the HPV method will not work. To solve this problem, the slope difference distribution (SDD) method was proposed [28] and it can be summarized as follows.

For the $i$th point of the smoothed histogram distribution $p'(x); x = 1,2,...,N$ , we compute two slopes for it by fitting two lines on its left side and on its right side respectively. The line equation is defined as follows.

$$l = ax + b \quad (40)$$

The coefficients $[a,b]^T$ of the line model are computed by the least squared fitting as follows.

$$[a,b]^T = (B^T B)^{-1} B^T Y \quad (41)$$

For the left fitted line, $B$ and $Y$ are formulated as follows.

$$B = \begin{bmatrix} i-Q+1 & 1 \\ i-Q+2 & 1 \\ \vdots & \vdots \\ i & 1 \end{bmatrix} \quad (42)$$

$$Y = \left[ p'(i-Q+1), p'(i-Q+2), \ldots, p'(i) \right]^T \quad (43)$$

Where $Q$ denotes the number of the selected points on the left of the $i$th point. For the right fitted line, $B$ and $Y$ are formulated by the following equation.

$$B = \begin{bmatrix} i & 1 \\ i+1 & 1 \\ \vdots & \vdots \\ i+Q-1 & 1 \end{bmatrix} \quad (44)$$

$$Y = \left[ p'(i), p'(i+1), \ldots, p'(i+Q-1) \right]^T \quad (45)$$

Where $Q$ denotes the number of the selected points on the right of the $i$th point. The slope difference at the $i$th point is then calculated by the following equation.

$$S_d(i) = a_{right}(i) - a_{left}(i); i = 1+Q,...,N-Q \quad (46)$$

Where $a_{right}(i)$ denotes the slope of the line fitted by the points selected on the right of the $i$th point and $a_{left}(i)$ denotes the slope of the line fitted by the points selected on the left side of the $i$th point. The derivative of the slope difference distribution $S_d(x)$ is computed and set to zero as follows.

$$\frac{dS_d(x)}{dx} = 0 \quad (47)$$

Both the peak feature points $P_k; k = 1,2,...,K$ and the valley feature points $V_k; k = 1,2,...,K-1$ of $S_d(x)$ can be calculated by solving the above equation. The means $\{\mu_1, \mu_2, ..., \mu_K\}$ are then calculated by Eq. (38) and the thresholds are then calculated by Eq. (39) or Eq. (7).

The synthesized image shown in Fig. 1 (a) is used to demonstrate the SDD method in Fig. 3 (b). Where "OH-D" denotes the original distribution of the histogram for the synthesized image, "FH-D" denotes the filtered distribution of the histogram formulated by Eq. (36) and "SD-D" denotes the distribution of the slope difference computed by Eq. (46). It is seen that the thresholds and the means calculated by the SDD method and by the HPV method are very similar for the histogram with the Gaussian mixture model distribution. However, the HPV method may not compute the thresholds or means as accurate as the SDD method when the histogram shapes become irregular, which will be demonstrated in the following sections.

## 6. Experimental Results

*6.1 Evaluation with the synthesized images*

In this section, the performances of the image segmentation methods based on histograms are evaluated quantitatively by segmenting the synthesized images since the histogram shapes and the parameters of the synthesized images could be controlled precisely. Three quantitative measures are computed to compare the accuracies of these methods. The first quantitative measure is called dice similarity coefficient (DSC) that is formulated as follows.

$$DSC = \frac{2 \times S \cap G}{S + G} \quad (48)$$

where $S$ denotes the computed area of the segmented object and $G$ denotes the ground truth area.

The second quantitative measure is called average perpendicular distance (APD) that is formulated as follows.

$$APD = \frac{1}{N_S} \sum_{i=1}^{N_S} \underset{j=1,2,...,N_G}{\arg\min} \left| P_i^S - P_j^G \right| \quad (49)$$

where $P_i^S$ represents the $i$th point of the segmented object's boundary and $P_j^G$ denotes the $j$th point on the groundtruth boundary. $N_S$ represents the total number of the points on the boundary of the segmented object and $N_G$ represents the total number of the points on the ground truth boundary.

The third quantitative measure is called average mean difference (AMD) that is formulated as follows.

$$AMD = \frac{1}{N_I \times N_\mu} \sum_{i=1}^{N_I} \sum_{j=1}^{N_\mu} \left| \mu_S(i,j) - \mu_G(i,j) \right| \qquad (50)$$

$\mu_S(i,j)$ represents the $j$th computed mean value for the $i$th image. $\mu_G(i,j)$ represents the $j$th ground-truth mean value in the $i$th image. $N_I$ represents the total number of the images used to compute the pixel means and $N_\mu$ represents the total number of the pixel means in each image.

Table 1. Comparison of the segmentation accuracies for different methods with the synthesized image sets {100,150,200; [11,90]}. (The bold values represent the best performances)

| Method | DSC | APD | AMD |
|---|---|---|---|
| DL-cellpose | 0.9695 | 1.5236 | NA |
| DL-MIL | 0.9618 | 1.8543 | NA |
| **DL-SAM** | **0.9816** | **0.9654** | NA |
| K-means | 0.8622 | 6.4877 | 41.6279 |
| Otsu | 0.9769 | 1.0299 | 3.8624 |
| Fuzzy-Cmeans | 0.9770 | 1.0313 | 3.9187 |
| EM-MAP | 0.8065 | 9.2061 | 42.1319 |
| EM-LSE | 0.8065 | 9.2004 | 45.4919 |
| EM-CC | 0.8064 | 9.1939 | 44.5058 |
| Max entropy | 0.8560 | 15.4130 | 132.3552 |
| Fuzzy entropy | 0.8496 | 16.1229 | 132.2897 |
| Cross entropy | 0.9766 | 1.0268 | 6.6674 |
| Metric cross entropy | 0.9766 | 1.0268 | 6.7041 |
| HPV | 0.9718 | 1.3637 | 3.4500 |
| SDD | 0.9765 | 1.0862 | **2.1875** |

The synthesized image is demonstrated in Fig. 1 (a). The mean value of the first pixel class, i.e., the background is 100. The mean value of the second pixel class, i.e., the two smaller rectangles is 150. The mean value of the third pixel class, i.e., the largest rectangle is 200. Then the Gaussian noises are added by multiplying the standard normal distribution with the amplitude varying from 11 to 90. The parameters of the synthesized image sets are denoted as {100,150,200; [11,90]} in this tutorial. The quantitative accuracies of these image segmentation methods based on histograms are compared in Table 1. In addition to comparing these histogram-based methods among themselves, we also compare them with three deep learning based image segmentation methods. The selected deep learning based image segmentation methods include DL-cellpose [35], the multiple instance learning (DL-MIL) framework [36] and the segment anything model (DL-SAM) [37]. It is seen that the DL-SAM method achieved the highest DSC score and the lowest APD error while SDD achieved the lowest AMD error. The K-means method, the EM methods, the max entropy method and the fuzzy entropy method achieved poor segmentation performances. To analyze the possible reasons that caused the poor performances of these methods, we plot the representative histograms of the synthesized images in Fig. 4 (a) and we plot their DSC scores in Fig. 4 (b). It is seen that the max entropy method performed badly for the first 15 images and the fuzzy entropy method performed badly for the first 16 images. The K-means method performed badly for the first 22 images and the EM methods performed badly for the first 32 images.

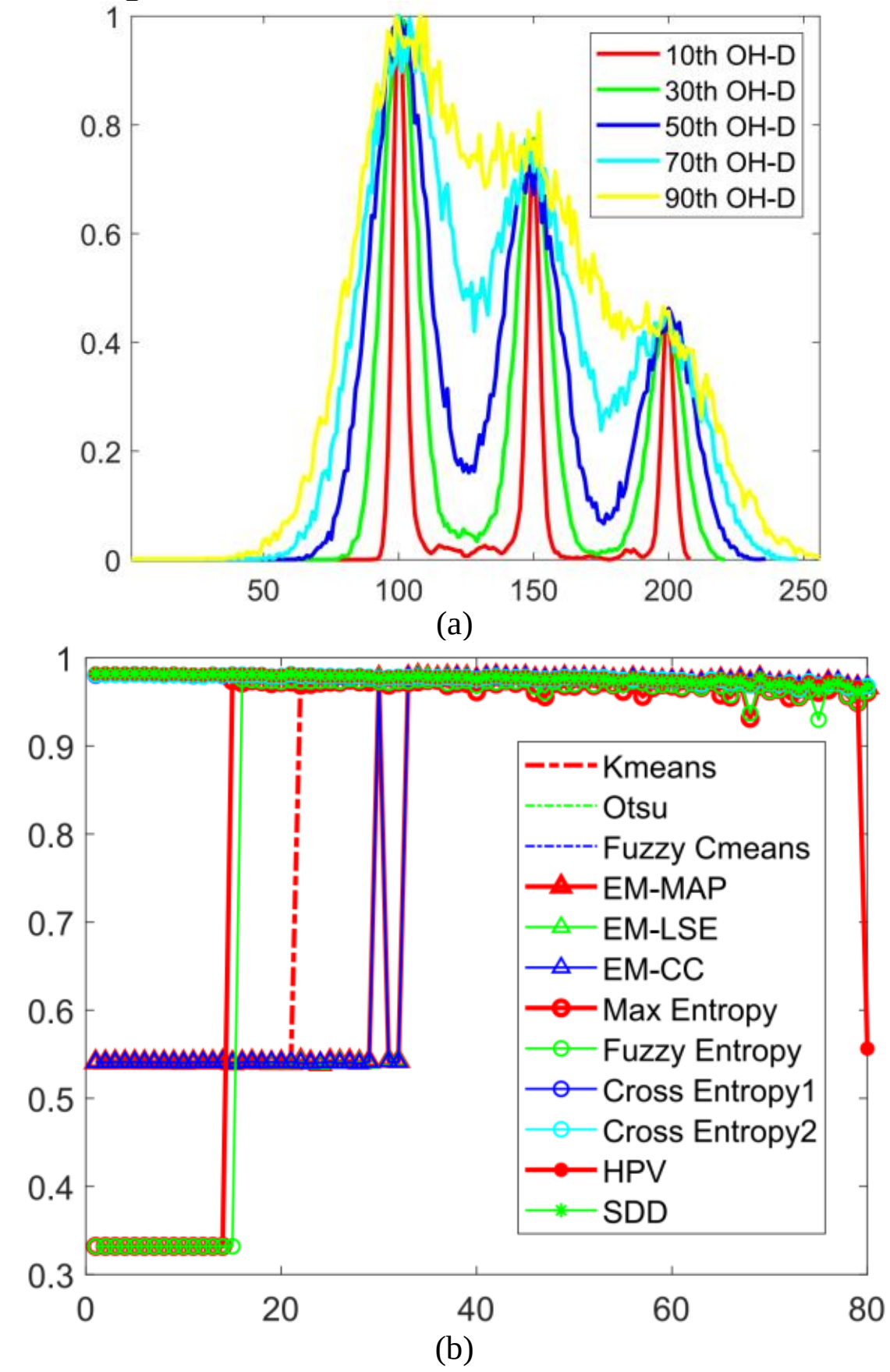

Fig. 4. Demonstration with the synthesized image sets {100,150,200; [11,90]}. (a) The histograms of the 10th synthesized image, the 30th synthesized image, the 50th synthesized image, the 70th synthesized image and the 90th synthesized image; (b) The computed DSC scores of the histogram-based image segmentation methods.

Table 2. Comparison of the segmentation accuracies for different methods with the synthesized image sets {50,120,200; [11,90]}. (The bold values represent the best performances)

| Method | DSC | APD | AMD |
|---|---|---|---|
| DL-cellpose | 0.9724 | 1.3760 | NA |
| DL-MIL | 0.9644 | 1.8231 | NA |
| **DL-SAM** | **0.9841** | **0.7072** | NA |
| K-means | 0.9799 | 0.7970 | 3.9048 |
| Otsu | 0.9799 | 0.7964 | 3.9261 |
| Fuzzy-Cmeans | 0.9799 | 0.7973 | 2.9940 |
| EM-MAP | 0.9798 | 0.8013 | 4.1658 |
| EM-LSE | 0.9798 | 0.8013 | 4.1678 |
| EM-CC | 0.9798 | 0.8020 | 4.3153 |
| Max entropy | 0.9756 | 1.1114 | 45.4902 |
| Fuzzy entropy | 0.9771 | 0.8360 | 29.0177 |
| Cross entropy | 0.9788 | 0.8002 | 5.2548 |
| Metric cross entropy | 0.9791 | 0.7940 | 5.1844 |
| HPV | 0.9802 | 0.7995 | **2.1000** |
| SDD | 0.9801 | 0.8039 | 2.2750 |

<

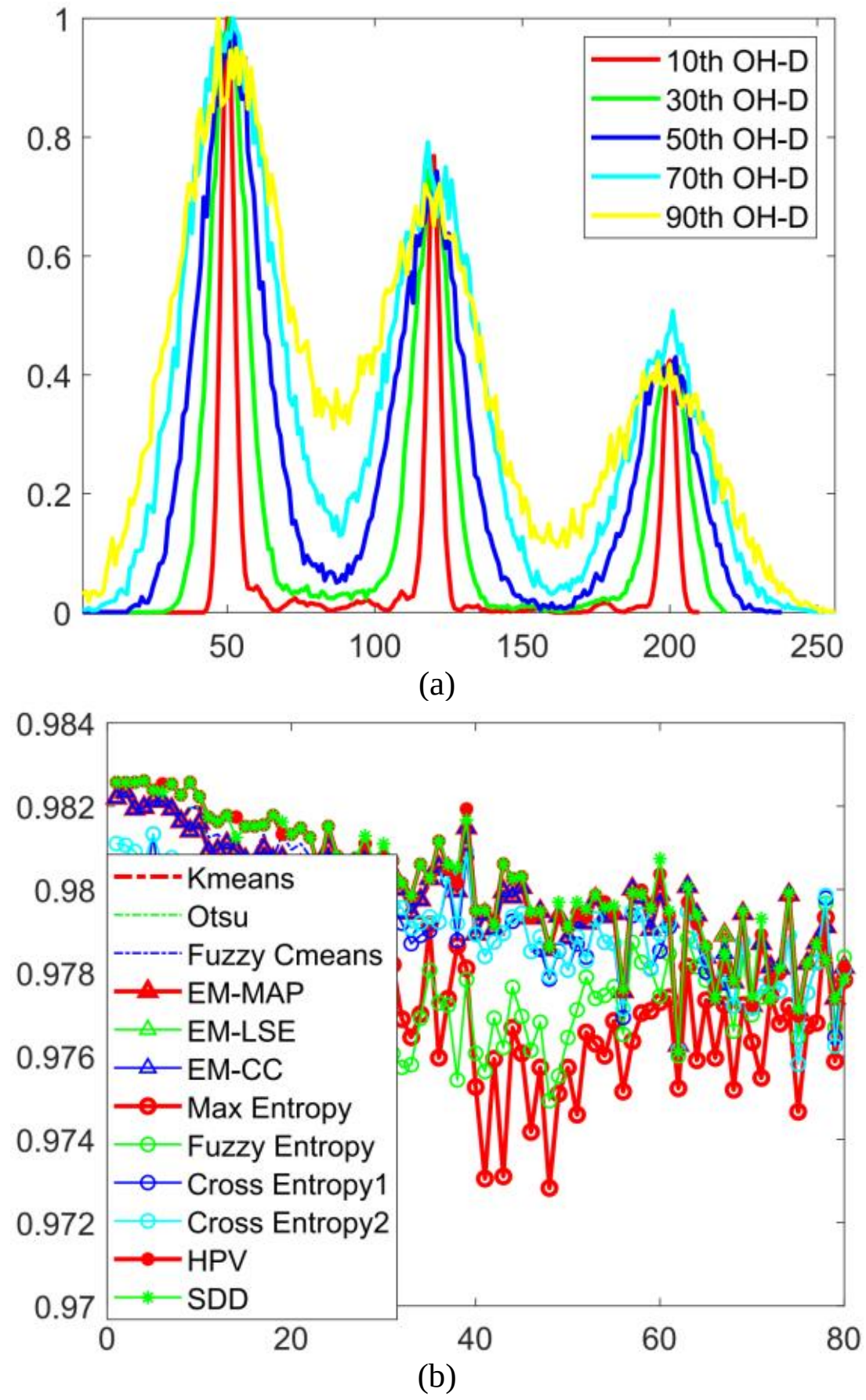

Fig. 5. Demonstration with the synthesized image sets {50,120,200; [11,90]}. (a) The histograms of the 10th synthesized image, the 30th synthesized image, the 50th synthesized image, the 70th synthesized image and the 90th synthesized image; (b) The computed DSC scores of the histogram-based image segmentation methods.

For comparison, we change the parameters of the synthesized image sets from {100,150,200; [11,90]} to {50,120,200; [11,90]}, i.e., we enlarge the distances between the adjacent pixel classes. The quantitative accuracies of these methods are shown in Table 2. It is seen that all these histogram-based image segmentation methods performed better and none of them performed badly anymore. To see the differences more clearly, we plot the representative histograms of the corresponding synthesized images in Fig. 5 (a) and we plot the DSC scores in Fig. 5 (b). It is seen that the histograms look similar except that the spacing between adjacent pixel classes becomes larger. The performances of most methods declined gradually with reference to the augments of the noise amplitude. Therefore, we could draw a conclusion that he performances of these methods are affected by the shapes of the histograms as demonstrated in Figs. 4-5.

We merge the background and the two smaller rectangles as the first pixel class and set their mean as 50. The pixels in the largest rectangle belong to the second pixel class and we set their mean as 200. The parameters of the synthesized image sets become {50,200; [11,90]}. The quantitative accuracies of these methods are shown in Table 3. It is seen that all these methods except the max entropy method performed well. If we change the parameter of the synthesized image sets to {150,200; [11,90]}, the K-means method, the EM methods, the max entropy method and the fuzzy entropy method will achieve poor segmentation performances, which corroborates the conclusion that the performances of the histogram-based methods are affected by the shapes of the histograms greatly. As can be seen from the quantitative comparisons shown in Tables 1-3, the DL-SAM method is more accurate than the histogram-based methods in segmenting the synthesized images while the other two deep learning methods are less accurate than the histogram-based methods.

Table 3. Comparison of the segmentation accuracies for different methods with the synthesized image sets {50,200; [11,90]}. (The bold values represent the best performances)

| Method | DSC | APD | AMD |
|---|---|---|---|
| DL-cellpose | 0.9736 | 1.3173 | NA |
| DL-MIL | 0.9654 | 1.7713 | NA |
| **DL-SAM** | **0.9908** | **0.1215** | NA |
| K-means | 0.9867 | 0.1668 | 2.1903 |
| Otsu | 0.9867 | 0.1665 | 2.1903 |
| Fuzzy-Cmeans | 0.9870 | 0.1484 | 1.6915 |
| EM-MAP | 0.9868 | 0.1598 | 2.3886 |
| EM-LSE | 0.9868 | 0.1597 | 2.3894 |
| EM-CC | 0.9868 | 0.1598 | 2.3892 |
| Max entropy | 0.8733 | 12.1279 | 8.5604 |
| Fuzzy entropy | 0.9392 | 4.7990 | 3.7257 |
| Cross entropy | 0.9825 | 0.3649 | 2.6484 |
| Metric cross entropy | 0.9842 | 0.2796 | 2.5467 |
| HPV | 0.9815 | 1.0076 | **0.5250** |
| SDD | 0.9871 | 0.1470 | 0.6500 |

Table 4. Comparison of the means of the computation time for different methods in segmenting the synthesized images with three pixel-classes and two pixel-classes respectively. (The bold values represent the best performances)

| Method | 3 means (millisecond) | 2 means (millisecond) |
|---|---|---|
| K-means | 0.52952 | 0.4145 |
| Otsu | 19.72625 | 0.5436 |
| Fuzzy-Cmeans | 1.09268 | 1.0274 |
| EM-MAP | 1.31419 | 0.4263 |
| EM-LSE | 1.00461 | 0.3731 |
| **EM-CC** | **0.46399** | **0.3535** |
| Max entropy | 618.60540 | 1.9638 |
| Fuzzy entropy | 1057.87737 | 1.9015 |
| Cross entropy | 779.31426 | 2.6434 |
| Metric cross entropy | 627.54222 | 1.6155 |
| HPV | 1.78593 | 0.6650 |
| SDD | 20.25115 | 5.6821 |

The mean computation time of each method is evaluated by segmenting 80 synthesized images of 3 means and 2 means respectively in MATLAB. The means of the computation time for these methods are shown in Table 4. The EM cross correlation method achieved the best performance and the fuzzy entropy method achieved the worst performance. Since the three compared deep-learning-based image segmentation methods are not implmented in MATLAB, their computation time are not available.

### 6.2 Evaluation with the real images

In this section, the performances of the image segmentation methods based on histograms are evaluated and compared with the deep learning methods quantitatively in segmenting the real images from different applications. The first image dataset is from cardiac left ventricle (LV) segmentation challenge [38] which is available online. 117 nuclear magnetic resonance imaging (MRI) images of the left ventricle at the end-diastolic stage or at the end-systolic stage are selected from the dataset. The quantitative accuracies of the image segmentation methods based on histograms and the image segmentation methods based on deep learning are compared in Table 5. It is seen that the SDD method achieved the best performance after the rules of the selecting the thresholds were designed according to the characteristics of the MRI images [29]. Most histogram-based image segmentation methods achieved better accuracies compared to the deep-learning methods [35-37] because these deep learning methods are not trained specially to segment this type of images.

Table 5. Comparison of the segmentation accuracies for different methods with the MICCAI MRI dataset. (The bold values represent the best performances)

| Method | DSC | APD |
|---|---|---|
| DL-cellpose | 0.8883 | 2.6939 |
| DL-MIL | 0.8853 | 2.7785 |
| DL-SAM | 0.8987 | 2.7668 |
| K-means | 0.9068 | 2.1845 |
| Otsu | 0.9073 | 2.1773 |
| Fuzzy-Cmeans | 0.9086 | 2.1473 |
| EM-MAP | 0.9046 | 2.3777 |
| EM-LSE | 0.9051 | 2.3523 |
| EM-CC | 0.9070 | 2.2037 |
| Max entropy | 0.9187 | 1.9206 |
| Fuzzy entropy | 0.9041 | 2.2339 |
| Cross entropy | 0.8929 | 2.7597 |
| Metric cross entropy | 0.8681 | 3.7335 |
| HPV | 0.8978 | 2.5144 |
| **SDD** | **0.9255** | **1.7700** |

The second image dataset is from the National University of Singapore (NUS) [39] which is available online. It is composed of 240 images of hand gestures in 10 classes. All the hand gesture images are with uniform background and they were captured by a CCD camera. The quantitative accuracies of different methods are compared in Table 6. It is seen that the K-means method and the Otsu method achieved the best DSC. The EM-MAP method achieved the best APD. The HPV method achieved the worst accuracy because there are no valleys in the smoothed histograms. The DL-cellpose method achieved the second worst accuracy because it was trained to segment simple shapes. Overall, the histogram-based image segmentation methods are more accurate than the deep-learning-based image segmentation methods [35-37] in this application because these deep learning methods are not trained specially to segment this type of images.

Table 6. Comparison of the segmentation accuracies for different methods with the NUS hand dataset I. (The bold values represent the best performances)

| Method | DSC | APD |
|---|---|---|
| DL-cellpose | 0.6255 | 50.5058 |
| DL-MIL | 0.9680 | 0.8109 |
| DL-SAM | 0.9885 | 0.2267 |
| **K-means** | **0.9891** | 0.4253 |
| **Otsu** | **0.9891** | 0.4253 |
| Fuzzy-Cmeans | 0.9883 | 0.2879 |
| EM-MAP | 0.9880 | **0.2107** |
| EM-LSE | 0.9787 | 0.3302 |
| EM-CC | 0.9787 | 0.3302 |
| Max entropy | 0.9617 | 2.3615 |
| Fuzzy entropy | 0.9787 | 0.3302 |
| Cross entropy | 0.9889 | 0.5908 |
| Metric cross entropy | 0.9889 | 0.5908 |
| HPV | 0.3255 | 80.0061 |
| SDD | 0.9860 | 0.510 |

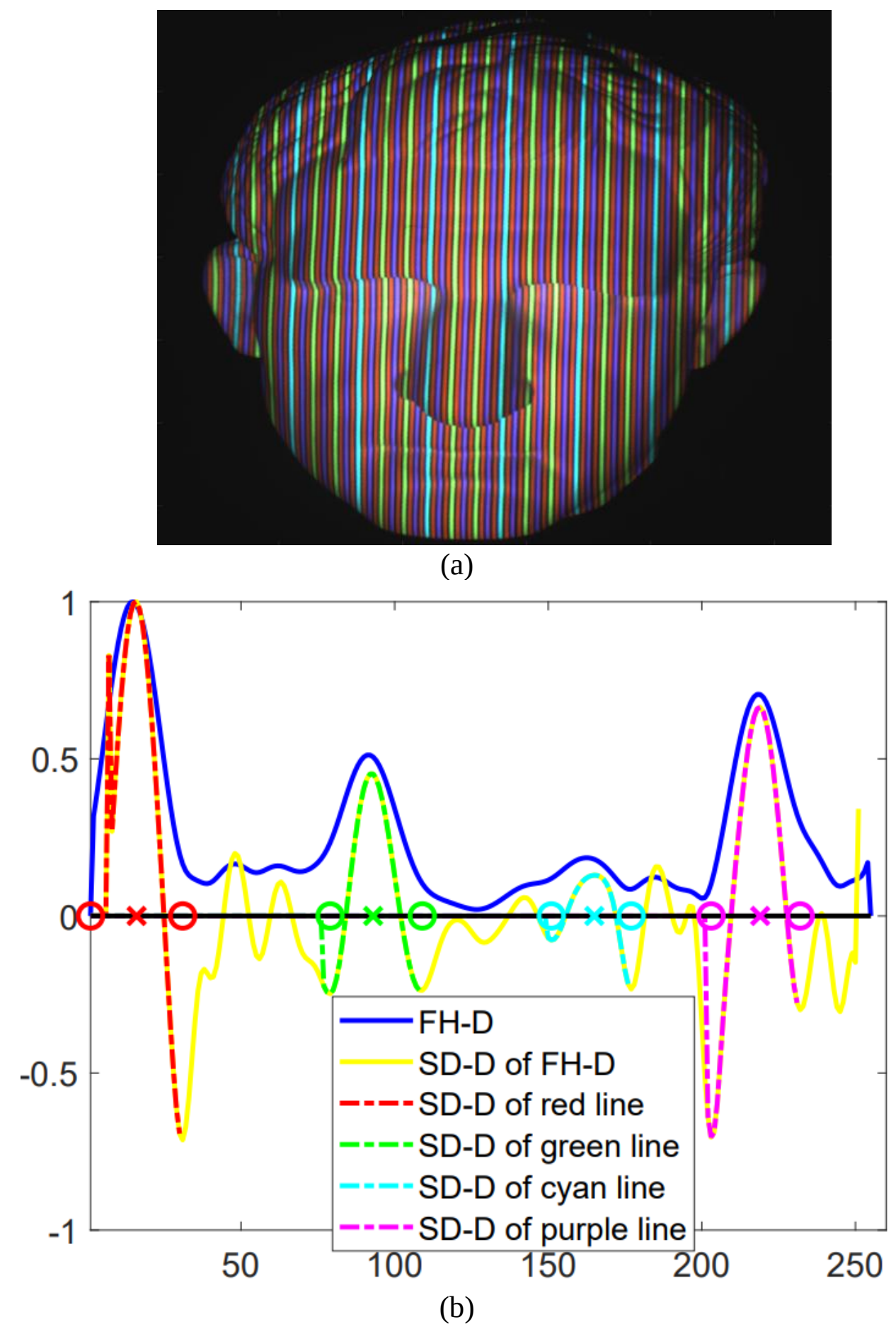

Fig. 6. Demonstration of the SDD method in segmenting a color line image. (a) A selected color line image; (b) The process of selecting the double SDD thresholds for different color lines.

<

The third image dataset is from our lab. It consists of 100 color line images. Different from the first two image datasets, multiple thresholds instead of one threshold are required to segment different color lines. The color line image contains the purple, red, cyan and red lines that need to be segmented and clustered for real-time 3D reconstruction. Since this image dataset is not available online, we demonstrate a typical color line image in Fig. 6 (a). The brightness of the color line image is eliminated by transforming the image from the RGB color space to the HSV color space. Then, the H channel is used to segment different color lines. The histogram of the color line image and its slope difference distribution are shown in Fig. 6 (b). It is seen that the pixel distribution of every type of color line follows the Gaussian distribution well. For the SDD method, the double thresholds are determined by the two valleys on both sides of the SDD peak computed for each type of color line. For other histogram-based methods, the double thresholds are determined by assigning a fixed range on both sides of the computed means. Based on trial and analysis, the fixed range is selected as [-15 15]. The quantitative accuracies of the histogram-based image segmentation methods are compared in Table 7 and we did not compare the deep-learning-based image segmentation methods because they could not generate meaningful segmentation results in this application. It is seen that the SDD method achieved the highest DSC scores. Both the SDD method and the HPV method perform better than other histogram-based methods in this application.

Table 7. Comparison of the segmentation accuracies for different methods with the color line image dataset. (The bold values represent the best performances)

| Method | DSC red lines | DSC green lines | DSC cyan lines | DSC purple lines | Average DSC |
|---|---|---|---|---|---|
| K-means | 0.9754 | 0.8723 | 0.9248 | 0.9401 | 0.9282 |
| Otsu | 0.8360 | 0.8467 | 0.9258 | 0.9392 | 0.8869 |
| Fuzzy-Cmeans | 0.9754 | 0.8723 | 0.9248 | 0.9401 | 0.9282 |
| EM-MAP | 0.9873 | 0.6980 | 0.9460 | 0.9466 | 0.8945 |
| EM-LSE | 0.9873 | 0.6979 | 0.9460 | 0.9466 | 0.8944 |
| EM-CC | 0.9873 | 0.6981 | 0.9459 | 0.9466 | 0.8945 |
| Max entropy | 0.0032 | 0.0000 | 0.5341 | 0.9187 | 0.3640 |
| Fuzzy entropy | 0.9720 | 0.8837 | 0.8890 | 0.9401 | 0.9212 |
| Cross entropy | 0.9809 | 0.8012 | 0.8839 | 0.9436 | 0.9024 |
| Metric cross entropy | 0.9800 | 0.8012 | 0.8870 | 0.9436 | 0.9030 |
| HPV | **0.9880** | 0.9699 | **0.9662** | 0.9678 | 0.9730 |
| **SDD** | **0.9880** | **0.9721** | 0.9576 | **0.9750** | **0.9732** |

The fourth image dataset is from the combined healthy abdominal organ segmentation (CHAOS) 2019 [40] which is available online. There are 20 batches of computed tomography (CT) liver images. One hundred images are selected to compare the image segmentation methods based on histograms and the image segmentation methods based on deep learning [35-37]. As it turned out, only the SDD method and the fuzzy C-means method could achieve meaningful segmentation results for the selected CT images. Two typical CT liver images are demonstrated in Figs. 7 (a)-(b) respectively. The ground truth contour of the liver delineated by the medical experts is represented by the red curve, the contour of the liver segmented by the SDD method is represented by the green curve and the contour of the liver segmented by the fuzzy C-means method is represented by the blue curve. Both methods compute two thresholds to segment the liver. The process of the compute the double thresholds by the SDD method is demontrated in Figs. 7 (c)-(d) respectively. It is seen that the distance between the liver pixel class and its adjacent pixel class is very small, which causes most the histogram-based image segmentation methods to treat them as one class. The double thresholds selected by the SDD method are denoted by the purple asterisks. The fuzzy C-means method also caculates two thresholds to segment the liver in the same way. The DSC computed by the SDD method on the selected 100 CT liver image is 0.9822 and the DSC computed by the fuzzy C-means method is 0.9824, which is very close. The APD computed by the SDD method on the selected 100 CT liver image is 1.5515 and the APD computed by the fuzzy C-means method is 1.5127, which is also very close. All the compared deep learning methods failed to yield meaningful results of the livers. We show the results of the DL-SAM method in Fig. 8 for comparison. It is seen that the DL-SAM method could not segment the livers robustly, which has aslo been reported in recent studies [4].

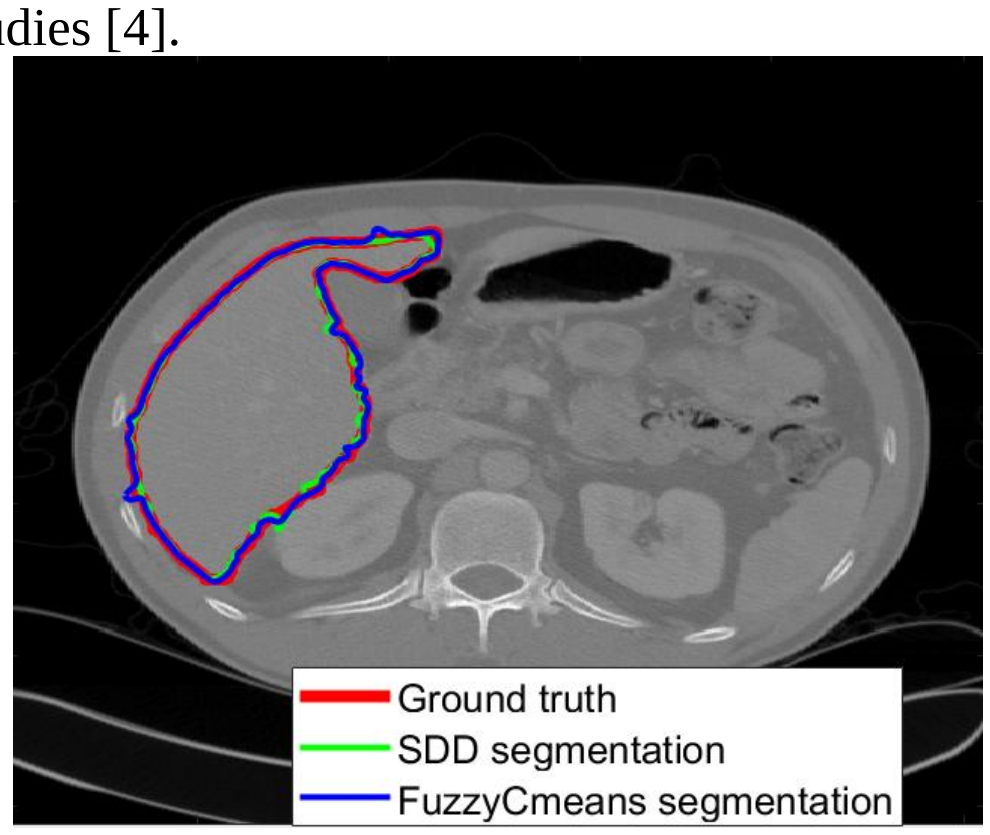

(a)

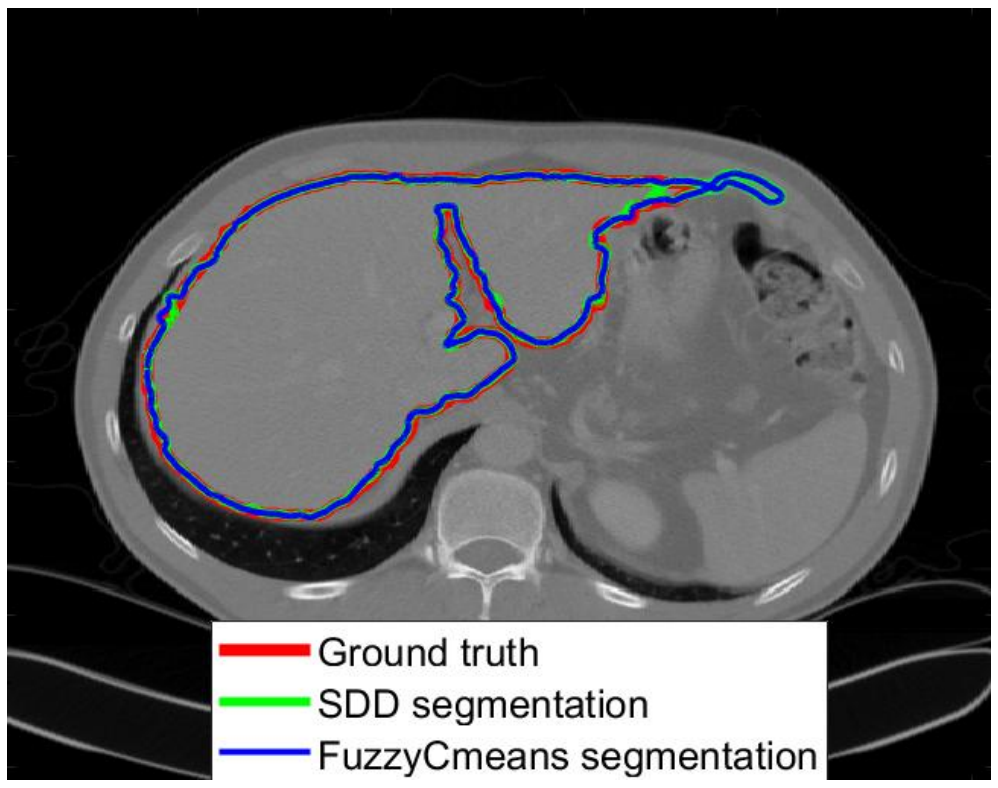

(b)

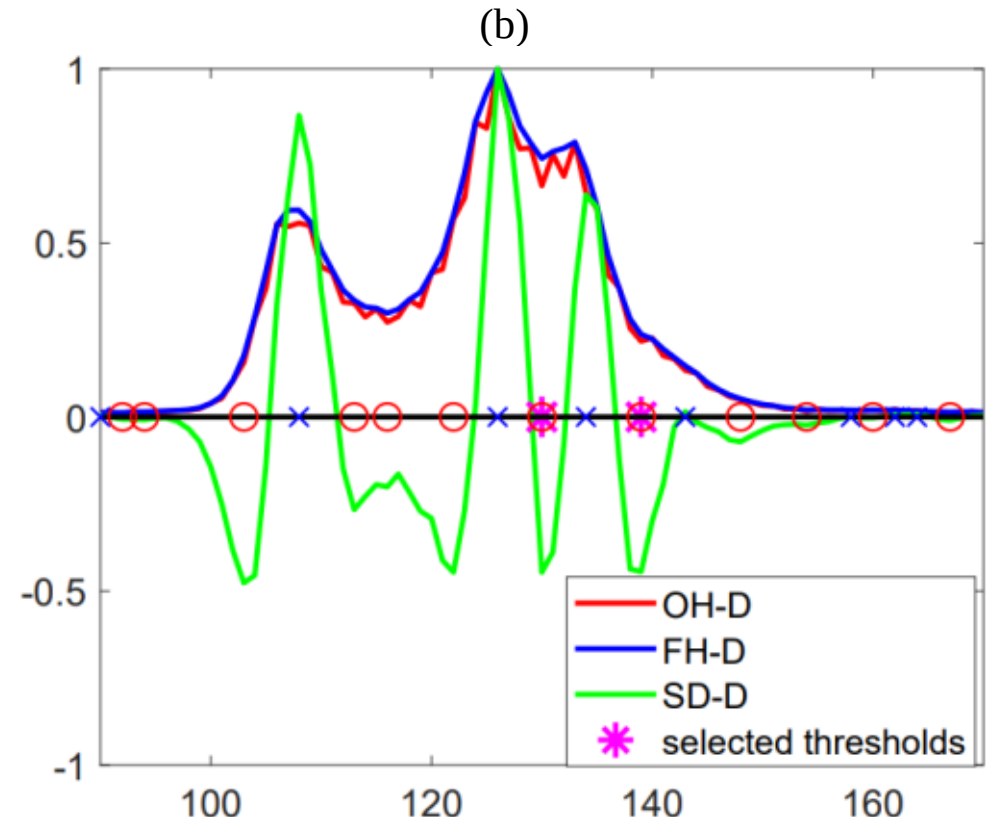

(c)

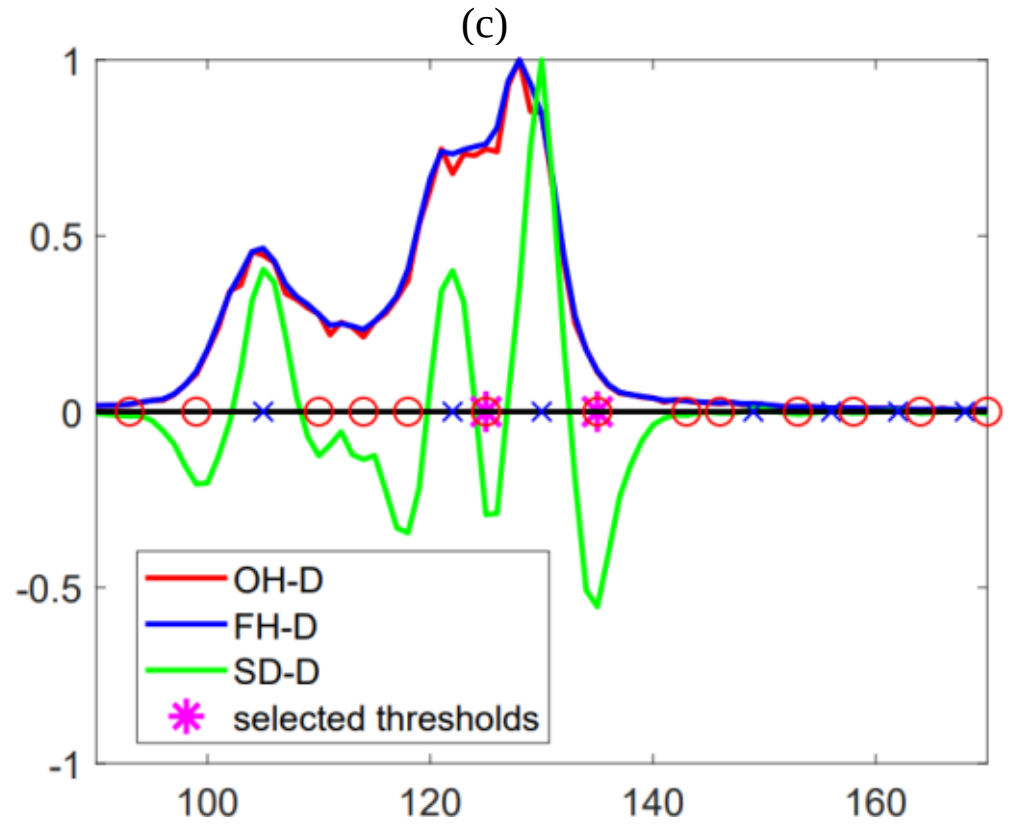

(d)

Fig. 7. Demonstration of the SDD method in segmenting the CT liver image. (a) A selected CT liver image with the ground-truth-contour and the segmented-contours- superimposed onto it; (b) Another selected CT liver image with the ground-truth-contour and the segmented-contours superimposed onto it; (c) The process of selecting the double SDD thresholds; (d) The process of selecting the double SDD thresholds.

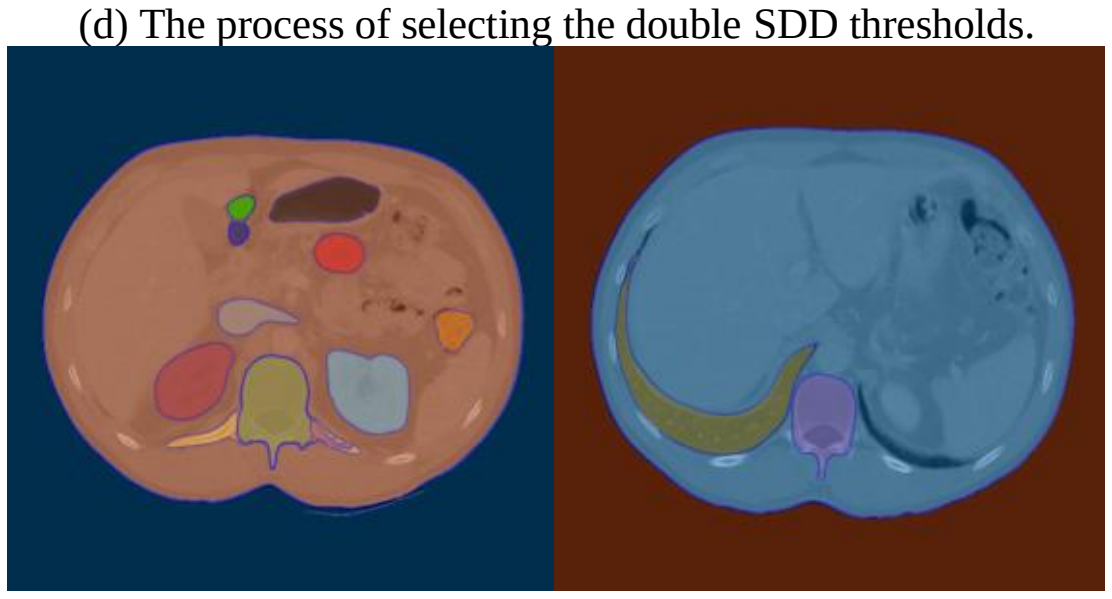

(a) (b)

Fig. 8. Demonstration of the DL-SAM method in segmenting the CT liver image. (a) Results of segmenting the first liver image; (b) Results of segmenting the second liver image.

## 7. Conclusion

This tutorial provided a short review of the histories and recent advances of the histogram-based image segmentation techniques. Their working principles were then described in a comparative way. Extensive experiments were conducted to compare the performances of the histogram-based image segmentation methods and three representative deep-learning-based image segmentation methods. The advantages and disadvantages of these techniques could be seen clearly with the extensive comparisons. The reason why so many image segmentation techniques have been developed so far is that there is no single technique that can solve the segmentation problem thoroughly. Based on the experimental results and comparisons in this tutorial, we could draw the following conclusions.

(1) there is not one single histogram-based image segmentation technique that can always achieve the best performance in all comparisons.

(2) the performances of the histogram-based image segmentation techniques are affected by the histogram shapes greatly. Among them, the SDD method and the fuzzy C-means method could adapt to a variety of histogram shapes while other histogram-based image segmentation techniques could not.

(3) the overall performance of the histogram-based image segmentation techniques is similar to the overall performance of the deep-learning-based image segmentation methods in the segmenting the synthesized images with uniform and simple backgrounds.

(4) without special training, the performances of the deep learning methods may be significantly inferior to the performances of the histogram-based image segmentation methods in the segmenting some types of complex images.